\documentclass[%
 reprint,
superscriptaddress,
 amsmath,amssymb,
 aps,
]{revtex4-2}
\usepackage{graphicx}
\usepackage{dcolumn}
\usepackage{bm}
\usepackage{color}
\usepackage{soul}
\usepackage[hidelinks]{hyperref}
\usepackage{natbib}

\begin{document}


\title{Controlled Propagation and Jamming of a Delamination Front}

\author{Mrityunjay Kothari}
\author{Zo\"e S. Lemon}
\affiliation{ Massachusetts Institute of Technology, Department of Civil and Environmental Engineering, Cambridge, MA, 02139, USA}
\author{Christine Roth}
\affiliation{ Massachusetts Institute of Technology, Department of Civil and Environmental Engineering, Cambridge, MA, 02139, USA}
\affiliation{Institute of Mechanical Engineering, École Polytechnique Fédérale de Lausanne (EPFL), 1015 Lausanne, Switzerland}
\author{Tal Cohen}
\email{Corresponding author: talco@mit.edu}
\affiliation{ Massachusetts Institute of Technology, Department of Civil and Environmental Engineering, Cambridge, MA, 02139, USA}
\affiliation{ Massachusetts Institute of Technology, Department of Mechanical Engineering, Cambridge, MA, 02139, USA}

\date{\today}

\begin{abstract}
We study the birth and propagation of a delamination front in the peeling of a soft, weakly adhesive layer. In a controlled-displacement setting, the layer partially detaches via a subcritical instability and the motion continues until arrested, by jamming of the two lobes. Using numerical solutions and scaling analysis, we quantitatively describe the equilibrium shapes and obtain constitutive sensitivities of jamming process to material and interface properties. We conclude with a way to delay or avoid jamming altogether by tunable interface properties. 


\end{abstract}
                              
\maketitle
 


\noindent The  crawling motion of a caterpillar serves as  a common pedagogical analogue to dislocation theory \citep{nabarro1989understanding,hirth1985brief}; although the soft bodied larva may not be able to push forward the entire length of its body simultaneously, without much effort it can form an arch at its hind, and work that arch forward to conquer a short distance. In a metal crystal, the caterpillar is representative of one sheet of atoms, and the arch, of one atomic spacing, which promotes deformation by moving. Clearly, these representations are oversimplified. On the one hand, the motion of an arch cannot accurately predict the behaviour in the bulk of a crystalline material. On the other hand, the locomotion of caterpillars is now known to be conspicuously more complex \citep{van2014locomotion}.   Regardless, studying and understanding physical phenomena that appear in the simplified  model  can ultimately feedback into our understanding of the analogous system. In this paper, we report on a jamming phenomenon that appears following the nucleation and propagation of a peeling arch in compliant, weakly adhesive layers. The arch  is formed by subjecting the layer to a controlled in-plane compressive displacement at one end.  Its motion is then resisted by the adhesive forces, until it is completely arrested by \textit{jamming}, thus producing stick-slip behaviour in a fully controlled manner. 

The motion of a ruck in a puckered carpet, serves as a similar and common analogue to dislocation theory and has been extensively studied in recent years \citep{kolinski2009shape,balmforth2015speed,lee2015role,vella2009statics}. There, the motion is resisted by gravitational forces. In contrast, here we consider the resistance due to adhesive forces between the layer and a substrate. This not only mimics the atomic interactions between  sheets, but also applies to a myriad  of additional physical systems: the stick-slip behavior of the delamination front is akin to a Shallamach wave \citep{schallamach1971does,rand2006insight,briggs1978rubber},
and is considered to be a fundamental mechanism in earthquakes \cite{brace1966stick,maegawa2010mechanism,ronsin2011nucleation}; evolution of geological formations also involves inter-layer binding forces  \cite{ramberg1964compression,schmalholz2002control}; adhesion and peeling are key mechanisms in emerging methods for fabrication and patterning of nanowires and flexible electronics \cite{shulaker2011linear,reis2009localization,vella2009macroscopic}; several additional examples appear in biology, where the control and propagation of adhesive interfaces is used to promote motility both at the level of a single cell \cite{schwarz2013physics,ziebert2016computational} and at the level of an entire organism \cite{trueman1975locomotion,cohen2018competing,labonte2016extreme}.

\begin{figure}[h]
	\centering
  \includegraphics[width = 0.38\textwidth]{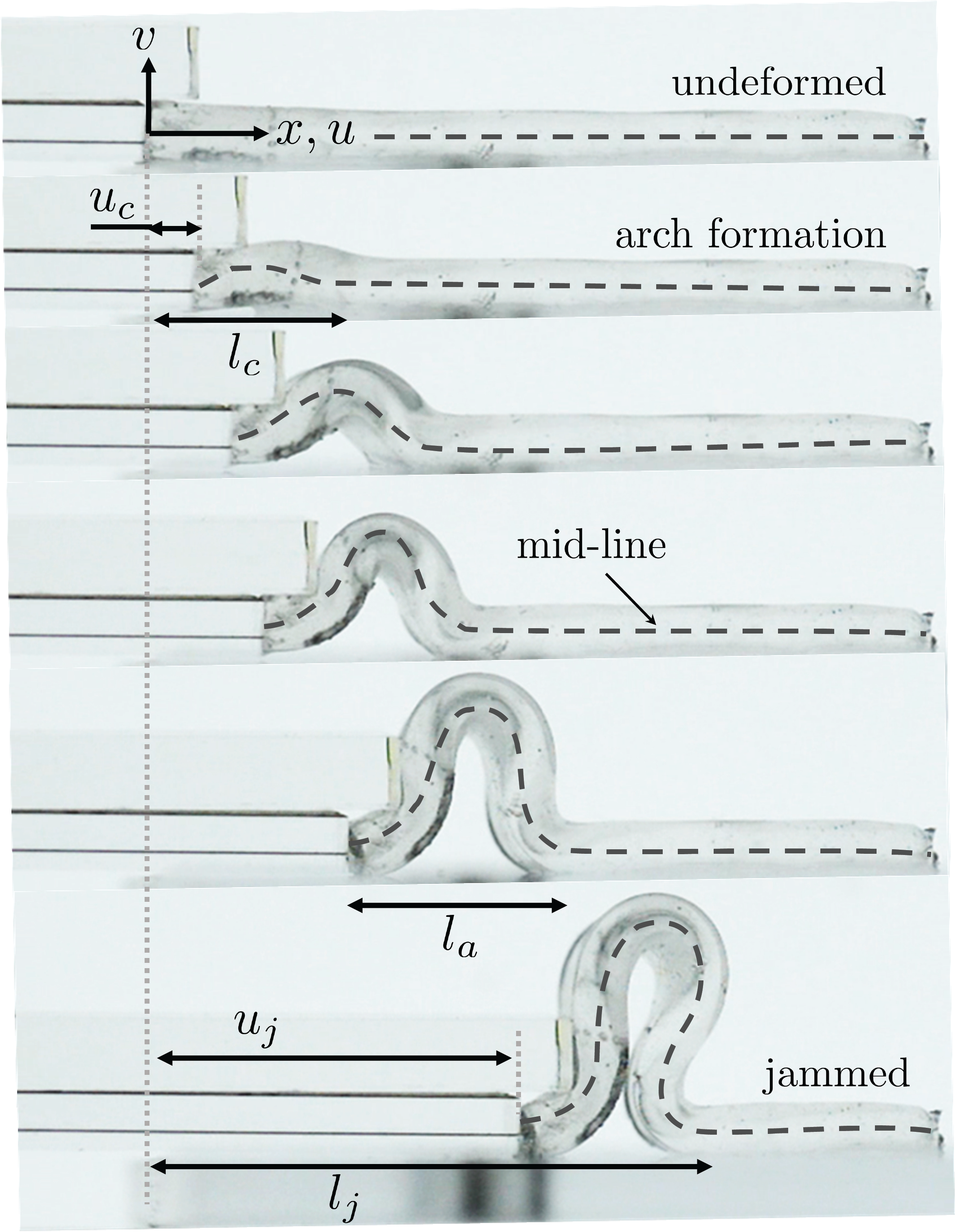} \vspace{-3mm}
  \caption{Controlled delamination in a PDMS layer of thickness $h=3.8$ mm and elastic modulus $E=140$ kPa \cite{SI}.}\label{fig1}
\end{figure}

  The delamination and jamming behavior described above is demonstrated by the model system in Fig.\ \ref{fig1}. A soft layer of PDMS is placed on a glass plate and compressed by controlling the horizontal displacement of its end $u_0$. The resulting horizontal and vertical displacements of the layer are denoted by $u(x)$ and $v(x)$, respectively, where   $x$, the material coordinate, is measured from the loaded end, such that $u_0=u(0)$.   Initially, the layer deforms in the plane while the adhesive bond appears to remain intact, until, at a critical displacement $u_c$, an arch forms thus delaminating a length  $l=l_c$ of the undeformed layer. As this quasistatic process continues, the arch length, $l_a=l-u_0$, of the delaminated region varies until the two sides of the arch come into contact, and are joined by the adhesive forces, thus arresting the peeling process. In the present formulation, all length scales are nondimensionalized with respect to the thickness of the layer, $h$.   
  
In this work, we facilitate this model system, which permits control over the entire delamination process, and employ an analytical model and scaling arguments to ask: \textit{How do the  properties of the layer and the adhesive interface influence the conditions for formation and  jamming of the delamination front? Can these material properties be tuned to allow the delamination front to propagate indefinitely?} 

A key property that must be accounted for, to answer these questions, is the nature of the adhesive interaction between the layer and the substrate. Considering small strains and weak bonding, such as the Van der Waals forces that attach  the elastomer layer to the glass in our model system (Fig.\ \ref{fig1}), we account for the effect of the  deformation of the adhesive bonds, prior to their detachment,  as previously suggested in \cite{cohen2018competing}. Accordingly, we divide the interface interactions into elastic and inelastic components. The bond stiffness, $k$, governs the elastic resistance to tangential (in-plane) sliding prior to detachment,  and the surface energy, $\Gamma$,  represents the  energy required to debond a unit area of the surface. 

For the mathematical derivation, we consider a  semi-infinite elastic layer that is bonded to the surface and occupies the region $0<x<\infty$ in its undeformed state. Upon displacement of the end by an amount $u_0$,  a section of length $l$ delaminates to form an arch of horizontal span $l_a=\int_0^l \cos\theta(x){\rm d}x$, where $\theta(x)$ is local angle between the arch tangent and the substrate, and changes in length of the  detached region are neglected \footnote{It has been verified that displacements due to compression of the delaminated region are negligible.}. The rest of the layer $(x>l)$ experiences compression $u'(x)<0$, but remains adhered to the surface. The displacement of the end is, thus, accommodated partly by formation of the arch and partly by the in-plane motion of the  adhered part with $u_l=u(l)=-\int_l^\infty u' {\rm d}x$. Accordingly, we write the kinematic constraint on the deformation $u_0=l-l_a+u_l$.
 
 For any prescribed $u_0$, the deformation of the layer, as determined by $\theta(x)$ and $u(x)$,  minimizes the total energy  of the system, per unit cross-sectional area (recall that all lengths are normalized with respect to  $h$) \vspace{-4mm}\begin{multline} \label{total_potential_energy}
    \Pi[\theta(x),u(x); l]= \int_{0}^{l}\bigg( \frac{EI}{2} \left(\frac{\theta'}{h}\right)^2  +2\Gamma \bigg){\rm d} x\\ +  \int_{l}^{\infty} \bigg(\frac{Eh}{2}(u')^2  + \frac{1}{2}k(hu)^2\bigg) {\rm d} x \\ +F\cdot\bigg(u_0 - \int_{0}^{l} (1-\cos\theta) {\rm d} x + \int_{l}^{\infty} u' {\rm d} x \bigg) 
\end{multline}
Here the first integral term corresponds to the delaminated part and comprises the elastic bending energy and the surface energy  invested in breaking the bonds, where $E$ is the Young's modulus, and $I$ is the area moment of inertia of a layer of unit width.
The second integral term corresponds to the adhered part, which stores energy due to elastic deformation of the layer and in extending the adhesive bonds, prior to their detachement.
In the last term of equation \eqref{total_potential_energy}, the displacement constraint is implemented by a Lagrange multiplier $F$ that can be interpreted as the   force  applied at $x=0$  to ensure the displacement $u_0$.

The optimal shape for an imposed displacement $u_0$ is found through a two step minimization procedure: first we find the optimal shape for a prescribed detached length $l$; then we find the minimum energy solution among  all values of $l$, including the fully adhered solution for which $l=0$. Mathematically we can write 
\vspace{-2mm}\begin{equation}\label{argmin}
    (\theta(x), u(x),l) ={\rm arg}~  \underset{l}{\rm min} \left( \underset{\theta, u}{\rm min } \ \Pi \right)
\end{equation}\vspace{-3mm}

The optimal solutions obtained in the first minimization follow from the Euler-Lagrange formulation, which results in two   differential equations, readily written in nondimensional form as\vspace{-2mm}
\begin{align}
\theta''-12f \sin\theta =0 & \quad\text{for} \quad x\in [0, l]  \label{ode1}\\
 \lambda^2u'' -u=0 & \quad\text{for} \quad x\in [ l ,\infty) \label{ode2}
\end{align}
where the dimensionless force  $f = F/Eh=-u'(l)$, derived using $I=h^3/12$, ensures balance of horizontal forces at $x=l$, and the dimensionless number $\lambda^2=E/kh$ emerges naturally from the formulation.


\begin{figure*}[!ht]
  \includegraphics[scale = 0.38]{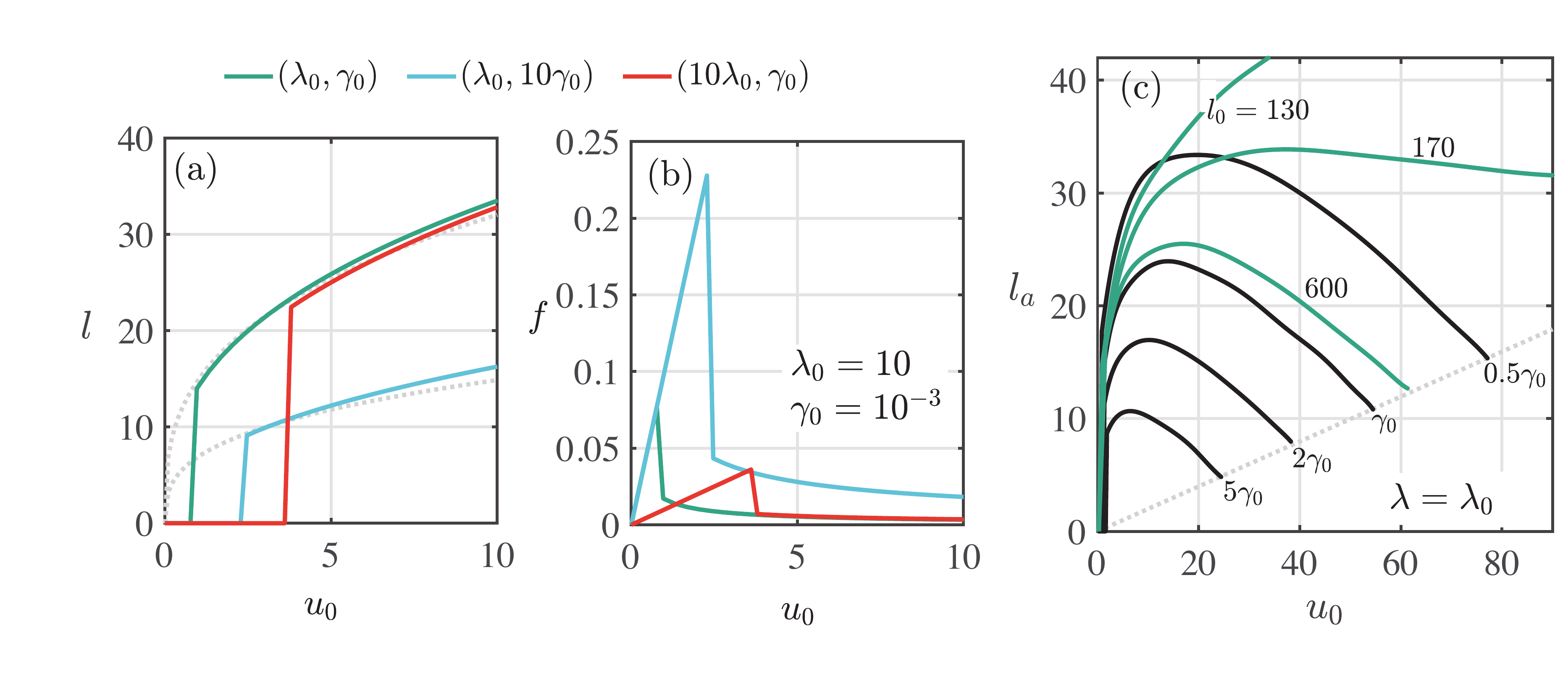}\vspace{-9mm}
  \caption{Theoretical results for dimensionless (a) detached length, (b) applied force, and (c) archlength, as functions of the applied displacement. In (c) colored curves correspond to layers with spatial variation of the surface energy $\gamma = \gamma_0e^{(-x/l_0)}$. By tuning the decay length $l_0$, it is possible to delay or even avoid jamming altogether.}\label{fig2}
\end{figure*}%

Equation (\ref{ode1}) is solved numerically, with zero angle maintained on both ends of the arch, namely $  \theta (x=0) = \theta(x=l) = 0$.
In the adhered region, the  displacement is expected to decay in the remote field, namely ${u(x \rightarrow \infty)} \rightarrow 0$, while its slope, at $x=l$, balances the force $f$.  Accordingly, we have $u(x)=\lambda f{\rm e}^{(l-x)/\lambda}$, with $v(x)\equiv 0$. From this result, it is apparent that $\lambda$ serves as a characteristic decay length in the adhered region. The dimensionless force, $f$, is obtained by the requirement of displacement continuity at $x=l$. 

For every prescribed displacement $u_0$, the above procedure is repeated for all values of $l$. The optimal $l$ that minimizes $\Pi$, is selected.

Representative results are shown in Fig.\ \ref{fig2}, where we examine the sensitivity of the process to the characteristic decay length $\lambda$ and the  dimensionless counterpart of the surface energy $\gamma=\Gamma/Eh$. The length of the delaminated region, $l$, is shown as a function of  $u_0$ in Fig.\ \ref{fig2}a. In all cases, delamination of a finite region appears as a first order transition, where the critical displacement, $u_c$, depends  on both $\lambda$ and $\gamma$. Quite interestingly, $\lambda$ influences the barrier for delamination, but has a negligible  effect on the shape of the curve that follows and, thus, on the shape of the arched region.    Prior to the delamination, the response is linear and is dictated solely by the elastic properties through $\lambda$, as shown by the force-displacement curves in Fig.\ \ref{fig2}b. Then, at $u_c$ a sudden drop in the applied force is observed and followed by a monotonic decrease  as displacement progresses. 

To study  the jamming limit, we examine  the evolution of the archlength as shown by the black curves in Fig.\ \ref{fig2}c, for different values of $\gamma$. In all cases, following the initial detachment, the arch increases its length up to a maximal value, beyond which the tendency is reversed and the arch begins to contract with increasing $u_0$, until it jams. This limit is estimated here by the intersection between the two sides of the layers' mid-line. Upon jamming, the shapes of the arched region in different layers are self-similar, namely jamming occurs once the ratio  $l_a/u_0\sim 0.2$ is obtained, as shown by the dashed grey line in Fig.\ \ref{fig2}c.

The sensitivities observed in Fig.\ \ref{fig2} can be further clarified by a scaling analysis. We consider arch formation under the assumption $u_0 \ll l$ and, thus, $\theta\ll 1$. At this limit,  equation \eqref{ode1} can be integrated analytically, which upon implementation  of  boundary conditions, reads $\theta(x)=A\sin{(2\pi x/l)}$, where  the compatibility requirement translates to 
$A^2=4{(u_0-u_l)/l}$, and force balance implies $u_l=\lambda f=\lambda\pi^2/3l^2$. Now, we substitute these relationships into \eqref{total_potential_energy} and omit small terms under the assumption $\lambda/l^4\ll 1$, to obtain the total energy $\Pi\sim Eh(\pi^2u_0/3l^2+2\gamma l)$, which admits a minimal value at\vspace{-5.5mm}
\begin{equation}\label{lsim}
 l\sim\left(\frac{\pi^2u_0}{3\gamma}\right)^{1/3}  \vspace{-2mm} 
\end{equation} Corresponding curves are shown by the dotted grey lines in Fig.\ \ref{fig2}a, and agree well with the numerical results for moderate displacements, $u_0$. 
\begin{figure}[ht]
	\centering
  \includegraphics[scale = 0.32]{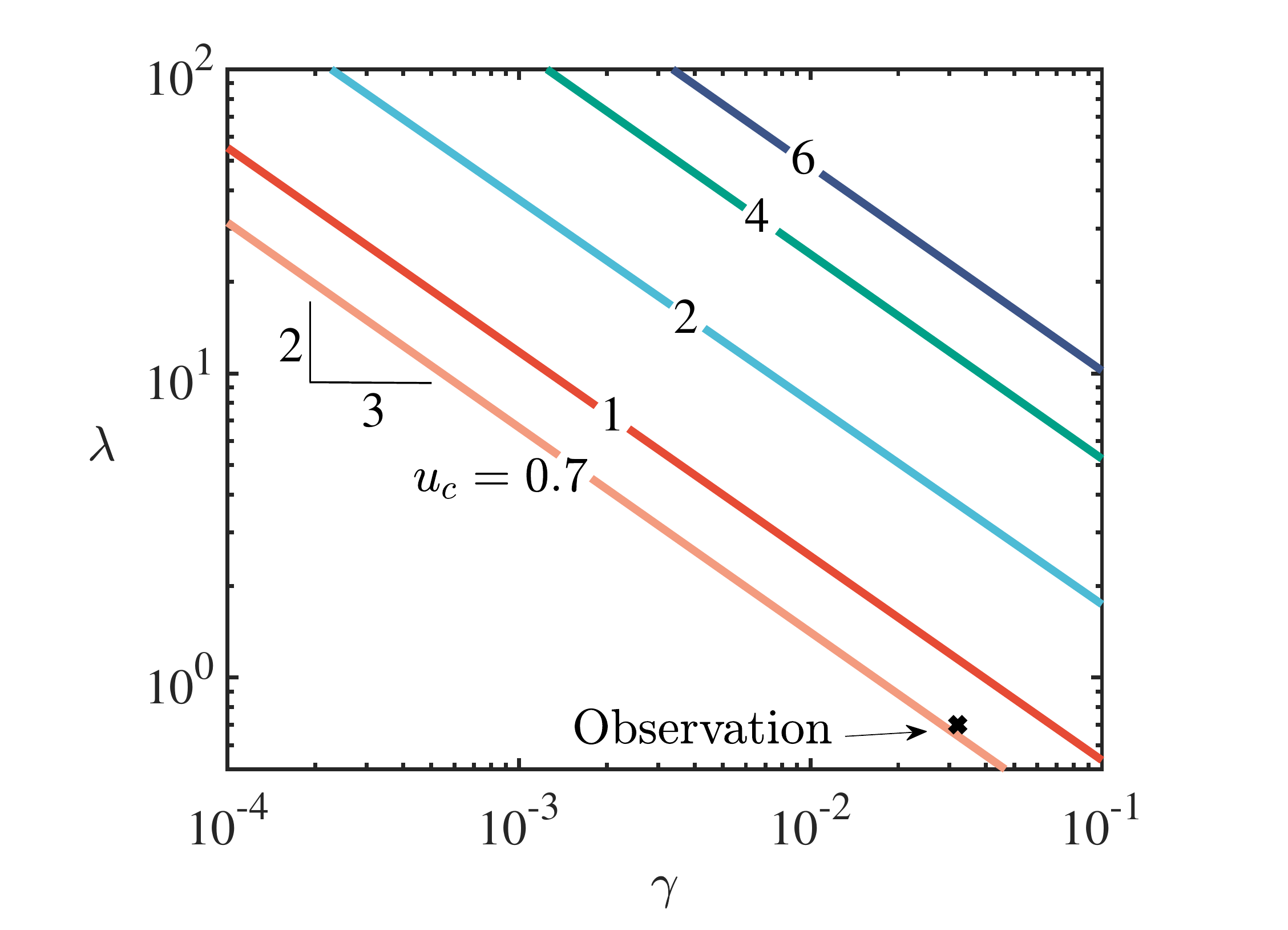}\vspace{-6mm}
  \caption{Phase portrait showing constant critical displacement curves on the $\gamma-\lambda$ plane.}\label{fig3}
\end{figure}%

Next, to estimate the critical values at which the arch forms, i.e. $(u_c,l_c)$, we compare the total energy invested in deforming the layer into the arched configuration, with that of the layer in a fully adhered state, i.e. $\Pi=Eh(u_0^2/2\lambda)$. Initially,  the flat configuration is energetically favorable. Then, the arch emerges when these energies intersect, which  after some algebra and by substitution of  \eqref{lsim} reads\vspace{-2mm}\begin{equation}\label{uc}
    u^5_c\sim72\pi^2\lambda^{3}\gamma^{2} \vspace{-2mm}
\end{equation}
We find a striking agreement between this relation and the results obtained via the numerical scheme, which are shown in the form of a phase portrait in Fig. \ref{fig3}. According to \eqref{uc}, at the limit of infinite bond stiffness  ($\lambda\to0$), which is commonly assumed in theories for interfacial fracture, an arch would form immediately upon loading with $u_c=0$, and would result in $l_c=0$, thus showing the importance of accounting for the bond stiffness to determine the nucleation limit. 

\begin{figure}[h]
	\centering
  \includegraphics[scale = 0.26]{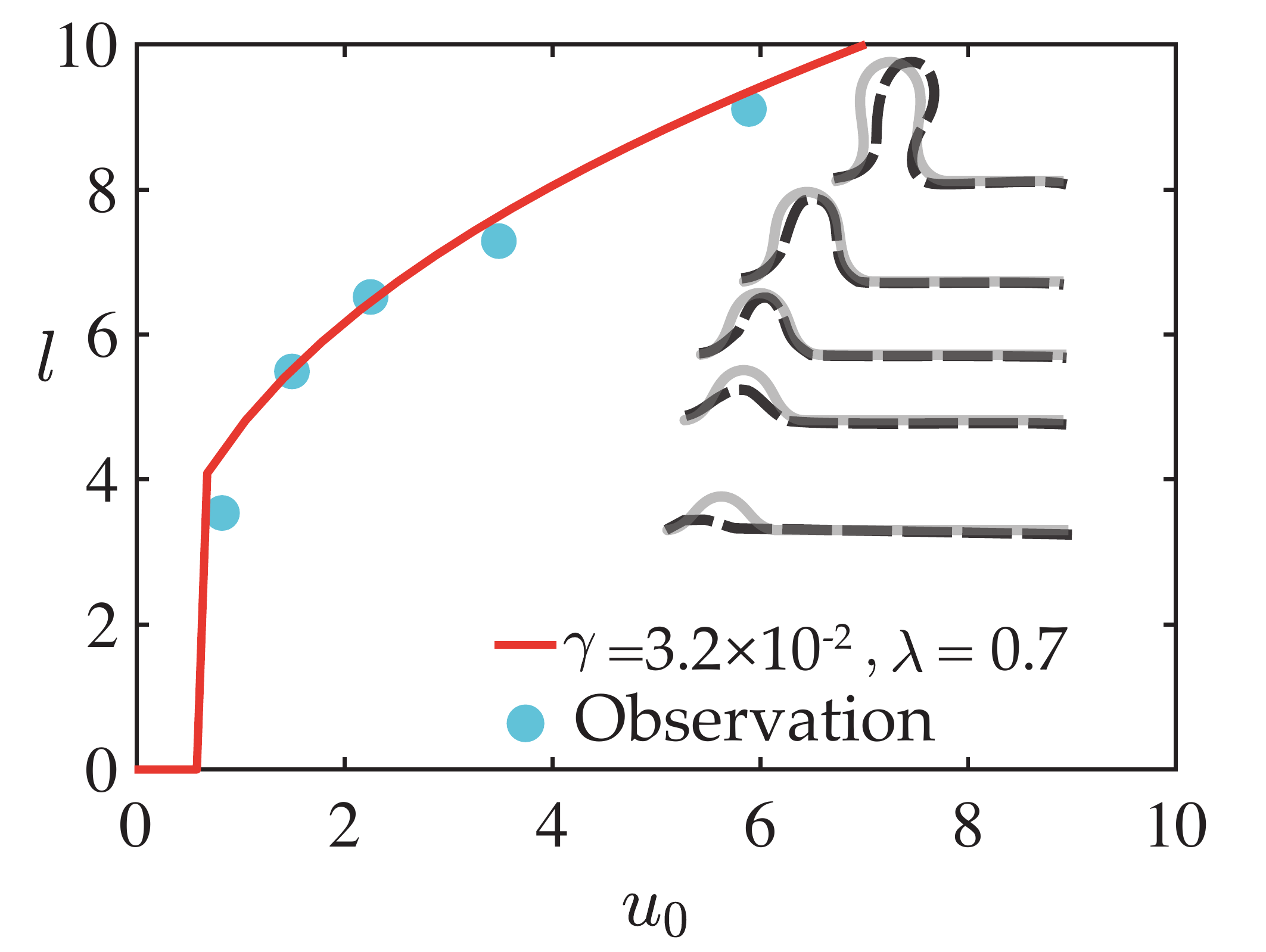}\vspace{-4mm}
  \caption{Comparison between observation and theory.}\label{fig4}
\end{figure}

To confirm the results of the above formulation, we now return to our observation in Fig. \ref{fig1}. We show the mid-line curves of the PDMS sample at different displacements in comparison with theoretical shapes, with $\lambda$ and $\gamma$ obtained by a best fit to the $l$ versus $u_0$ response, in Fig. \ref{fig4}.  The discrepancy in the arch shape  for small $u_0$, where the arch is shallow, is due to the extensibility of the layer. This effect becomes less significant as $u_0$ increases. At larger displacements, when jamming is approached, symmetry breakage may occur due to gravity, which is neglected in the present model.   From this comparison we have for the PDMS layer  $\gamma\sim3.2\times10^{-2}$ and $\lambda\sim 0.7$. This translates to the dimensional values $\Gamma\sim1.7\times10^{-2} $N/mm  and $k\sim0.075$N/mm$^3$, which agree with representative values in the literature \cite{sofla2010pdms,cohen2018competing}. The present peeling method thus provides a novel technique to measure these properties, which are otherwise difficult to obtain \footnote{Note that in PDMS surface properties vary significantly with stiffness.}. 

Finally, a key result of this work is the identification of the terminal jammed state and its dependence on the layer properties. While intermediate configurations are unstable to perturbations of the applied force, the jammed state is expected to be most ubiquitous in the natural world. Moreover, recent works have demonstrated that by employing surface wrinkling and kirigami techniques to control interface morphology, it is possible to tune the surface adhesive properties \cite{lin2008mechanically, hwang2018kirigami}. Therefore, given our understanding of this phenomenon, it is now possible to tune the system  not only to prescribe the jamming distance, but to avoid jamming completely. As an example, the latter can be achieved by  spatially varying the surface energy, as shown in Fig. \ref{fig2}c. This potential of active control raises the question as to whether caterpillars or even cells exploit similar in-plane mechanisms to form an arch and control their motion, and can artificial smart systems do the same?

%


\begin{thebibliography}{29}%
\makeatletter
\providecommand \@ifxundefined [1]{%
 \@ifx{#1\undefined}
}%
\providecommand \@ifnum [1]{%
 \ifnum #1\expandafter \@firstoftwo
 \else \expandafter \@secondoftwo
 \fi
}%
\providecommand \@ifx [1]{%
 \ifx #1\expandafter \@firstoftwo
 \else \expandafter \@secondoftwo
 \fi
}%
\providecommand \natexlab [1]{#1}%
\providecommand \enquote  [1]{``#1''}%
\providecommand \bibnamefont  [1]{#1}%
\providecommand \bibfnamefont [1]{#1}%
\providecommand \citenamefont [1]{#1}%
\providecommand \href@noop [0]{\@secondoftwo}%
\providecommand \href [0]{\begingroup \@sanitize@url \@href}%
\providecommand \@href[1]{\@@startlink{#1}\@@href}%
\providecommand \@@href[1]{\endgroup#1\@@endlink}%
\providecommand \@sanitize@url [0]{\catcode `\\12\catcode `\$12\catcode
  `\&12\catcode `\#12\catcode `\^12\catcode `\_12\catcode `\%12\relax}%
\providecommand \@@startlink[1]{}%
\providecommand \@@endlink[0]{}%
\providecommand \url  [0]{\begingroup\@sanitize@url \@url }%
\providecommand \@url [1]{\endgroup\@href {#1}{\urlprefix }}%
\providecommand \urlprefix  [0]{URL }%
\providecommand \Eprint [0]{\href }%
\providecommand \doibase [0]{https://doi.org/}%
\providecommand \selectlanguage [0]{\@gobble}%
\providecommand \bibinfo  [0]{\@secondoftwo}%
\providecommand \bibfield  [0]{\@secondoftwo}%
\providecommand \translation [1]{[#1]}%
\providecommand \BibitemOpen [0]{}%
\providecommand \bibitemStop [0]{}%
\providecommand \bibitemNoStop [0]{.\EOS\space}%
\providecommand \EOS [0]{\spacefactor3000\relax}%
\providecommand \BibitemShut  [1]{\csname bibitem#1\endcsname}%
\let\auto@bib@innerbib\@empty
\bibitem [{\citenamefont {Nabarro}(1989)}]{nabarro1989understanding}%
  \BibitemOpen
  \bibfield  {author} {\bibinfo {author} {\bibfnamefont {F.}~\bibnamefont
  {Nabarro}},\ }\bibfield  {title} {\bibinfo {title} {Understanding the
  mechanical properties of metals and alioys (765kb)},\ }\href@noop {}
  {\bibfield  {journal} {\bibinfo  {journal} {South African Journal of
  Science}\ }\textbf {\bibinfo {volume} {85}},\ \bibinfo {pages} {589}
  (\bibinfo {year} {1989})}\BibitemShut {NoStop}%
\bibitem [{\citenamefont {Hirth}(1985)}]{hirth1985brief}%
  \BibitemOpen
  \bibfield  {author} {\bibinfo {author} {\bibfnamefont {J.}~\bibnamefont
  {Hirth}},\ }\bibfield  {title} {\bibinfo {title} {A brief history of
  dislocation theory},\ }\href@noop {} {\bibfield  {journal} {\bibinfo
  {journal} {Metallurgical Transactions A}\ }\textbf {\bibinfo {volume} {16}},\
  \bibinfo {pages} {2085} (\bibinfo {year} {1985})}\BibitemShut {NoStop}%
\bibitem [{\citenamefont {Van~Griethuijsen}\ and\ \citenamefont
  {Trimmer}(2014)}]{van2014locomotion}%
  \BibitemOpen
  \bibfield  {author} {\bibinfo {author} {\bibfnamefont {L.}~\bibnamefont
  {Van~Griethuijsen}}\ and\ \bibinfo {author} {\bibfnamefont {B.}~\bibnamefont
  {Trimmer}},\ }\bibfield  {title} {\bibinfo {title} {Locomotion in
  caterpillars},\ }\href@noop {} {\bibfield  {journal} {\bibinfo  {journal}
  {Biological Reviews}\ }\textbf {\bibinfo {volume} {89}},\ \bibinfo {pages}
  {656} (\bibinfo {year} {2014})}\BibitemShut {NoStop}%
\bibitem [{\citenamefont {Kolinski}\ \emph {et~al.}(2009)\citenamefont
  {Kolinski}, \citenamefont {Aussillous},\ and\ \citenamefont
  {Mahadevan}}]{kolinski2009shape}%
  \BibitemOpen
  \bibfield  {author} {\bibinfo {author} {\bibfnamefont {J.~M.}\ \bibnamefont
  {Kolinski}}, \bibinfo {author} {\bibfnamefont {P.}~\bibnamefont
  {Aussillous}},\ and\ \bibinfo {author} {\bibfnamefont {L.}~\bibnamefont
  {Mahadevan}},\ }\bibfield  {title} {\bibinfo {title} {Shape and motion of a
  ruck in a rug},\ }\href@noop {} {\bibfield  {journal} {\bibinfo  {journal}
  {Physical review letters}\ }\textbf {\bibinfo {volume} {103}},\ \bibinfo
  {pages} {174302} (\bibinfo {year} {2009})}\BibitemShut {NoStop}%
\bibitem [{\citenamefont {Balmforth}\ \emph {et~al.}(2015)\citenamefont
  {Balmforth}, \citenamefont {Craster},\ and\ \citenamefont
  {Hewitt}}]{balmforth2015speed}%
  \BibitemOpen
  \bibfield  {author} {\bibinfo {author} {\bibfnamefont {N.}~\bibnamefont
  {Balmforth}}, \bibinfo {author} {\bibfnamefont {R.}~\bibnamefont {Craster}},\
  and\ \bibinfo {author} {\bibfnamefont {I.}~\bibnamefont {Hewitt}},\
  }\bibfield  {title} {\bibinfo {title} {The speed of an inclined ruck},\
  }\href@noop {} {\bibfield  {journal} {\bibinfo  {journal} {Proceedings of the
  Royal Society A: Mathematical, Physical and Engineering Sciences}\ }\textbf
  {\bibinfo {volume} {471}},\ \bibinfo {pages} {20140740} (\bibinfo {year}
  {2015})}\BibitemShut {NoStop}%
\bibitem [{\citenamefont {Lee}\ \emph {et~al.}(2015)\citenamefont {Lee},
  \citenamefont {Le~Gouellec},\ and\ \citenamefont {Vella}}]{lee2015role}%
  \BibitemOpen
  \bibfield  {author} {\bibinfo {author} {\bibfnamefont {A.~A.}\ \bibnamefont
  {Lee}}, \bibinfo {author} {\bibfnamefont {C.}~\bibnamefont {Le~Gouellec}},\
  and\ \bibinfo {author} {\bibfnamefont {D.}~\bibnamefont {Vella}},\ }\bibfield
   {title} {\bibinfo {title} {The role of extensibility in the birth of a ruck
  in a rug},\ }\href@noop {} {\bibfield  {journal} {\bibinfo  {journal}
  {Extreme Mechanics Letters}\ }\textbf {\bibinfo {volume} {5}},\ \bibinfo
  {pages} {81} (\bibinfo {year} {2015})}\BibitemShut {NoStop}%
\bibitem [{\citenamefont {Vella}\ \emph
  {et~al.}(2009{\natexlab{a}})\citenamefont {Vella}, \citenamefont {Boudaoud},\
  and\ \citenamefont {Adda-Bedia}}]{vella2009statics}%
  \BibitemOpen
  \bibfield  {author} {\bibinfo {author} {\bibfnamefont {D.}~\bibnamefont
  {Vella}}, \bibinfo {author} {\bibfnamefont {A.}~\bibnamefont {Boudaoud}},\
  and\ \bibinfo {author} {\bibfnamefont {M.}~\bibnamefont {Adda-Bedia}},\
  }\bibfield  {title} {\bibinfo {title} {Statics and inertial dynamics of a
  ruck in a rug},\ }\href@noop {} {\bibfield  {journal} {\bibinfo  {journal}
  {Physical review letters}\ }\textbf {\bibinfo {volume} {103}},\ \bibinfo
  {pages} {174301} (\bibinfo {year} {2009}{\natexlab{a}})}\BibitemShut
  {NoStop}%
\bibitem [{\citenamefont {Schallamach}(1971)}]{schallamach1971does}%
  \BibitemOpen
  \bibfield  {author} {\bibinfo {author} {\bibfnamefont {A.}~\bibnamefont
  {Schallamach}},\ }\bibfield  {title} {\bibinfo {title} {How does rubber
  slide?},\ }\href@noop {} {\bibfield  {journal} {\bibinfo  {journal} {Wear}\
  }\textbf {\bibinfo {volume} {17}},\ \bibinfo {pages} {301} (\bibinfo {year}
  {1971})}\BibitemShut {NoStop}%
\bibitem [{\citenamefont {Rand}\ and\ \citenamefont
  {Crosby}(2006)}]{rand2006insight}%
  \BibitemOpen
  \bibfield  {author} {\bibinfo {author} {\bibfnamefont {C.~J.}\ \bibnamefont
  {Rand}}\ and\ \bibinfo {author} {\bibfnamefont {A.~J.}\ \bibnamefont
  {Crosby}},\ }\bibfield  {title} {\bibinfo {title} {Insight into the
  periodicity of schallamach waves in soft material friction},\ }\href@noop {}
  {\bibfield  {journal} {\bibinfo  {journal} {Applied physics letters}\
  }\textbf {\bibinfo {volume} {89}},\ \bibinfo {pages} {261907} (\bibinfo
  {year} {2006})}\BibitemShut {NoStop}%
\bibitem [{\citenamefont {Briggs}\ and\ \citenamefont
  {Briscoe}(1978)}]{briggs1978rubber}%
  \BibitemOpen
  \bibfield  {author} {\bibinfo {author} {\bibfnamefont {G.}~\bibnamefont
  {Briggs}}\ and\ \bibinfo {author} {\bibfnamefont {B.}~\bibnamefont
  {Briscoe}},\ }\bibfield  {title} {\bibinfo {title} {How rubber grips and
  slips schallamach waves and the friction of elastomers},\ }\href@noop {}
  {\bibfield  {journal} {\bibinfo  {journal} {Philosophical Magazine A}\
  }\textbf {\bibinfo {volume} {38}},\ \bibinfo {pages} {387} (\bibinfo {year}
  {1978})}\BibitemShut {NoStop}%
\bibitem [{\citenamefont {Brace}\ and\ \citenamefont
  {Byerlee}(1966)}]{brace1966stick}%
  \BibitemOpen
  \bibfield  {author} {\bibinfo {author} {\bibfnamefont {W.}~\bibnamefont
  {Brace}}\ and\ \bibinfo {author} {\bibfnamefont {J.}~\bibnamefont
  {Byerlee}},\ }\bibfield  {title} {\bibinfo {title} {Stick-slip as a mechanism
  for earthquakes},\ }\href@noop {} {\bibfield  {journal} {\bibinfo  {journal}
  {Science}\ }\textbf {\bibinfo {volume} {153}},\ \bibinfo {pages} {990}
  (\bibinfo {year} {1966})}\BibitemShut {NoStop}%
\bibitem [{\citenamefont {Maegawa}\ and\ \citenamefont
  {Nakano}(2010)}]{maegawa2010mechanism}%
  \BibitemOpen
  \bibfield  {author} {\bibinfo {author} {\bibfnamefont {S.}~\bibnamefont
  {Maegawa}}\ and\ \bibinfo {author} {\bibfnamefont {K.}~\bibnamefont
  {Nakano}},\ }\bibfield  {title} {\bibinfo {title} {Mechanism of stick-slip
  associated with schallamach waves},\ }\href@noop {} {\bibfield  {journal}
  {\bibinfo  {journal} {Wear}\ }\textbf {\bibinfo {volume} {268}},\ \bibinfo
  {pages} {924} (\bibinfo {year} {2010})}\BibitemShut {NoStop}%
\bibitem [{\citenamefont {Ronsin}\ \emph {et~al.}(2011)\citenamefont {Ronsin},
  \citenamefont {Baumberger},\ and\ \citenamefont
  {Hui}}]{ronsin2011nucleation}%
  \BibitemOpen
  \bibfield  {author} {\bibinfo {author} {\bibfnamefont {O.}~\bibnamefont
  {Ronsin}}, \bibinfo {author} {\bibfnamefont {T.}~\bibnamefont {Baumberger}},\
  and\ \bibinfo {author} {\bibfnamefont {C.}~\bibnamefont {Hui}},\ }\bibfield
  {title} {\bibinfo {title} {Nucleation and propagation of quasi-static
  interfacial slip pulses},\ }\href@noop {} {\bibfield  {journal} {\bibinfo
  {journal} {The Journal of Adhesion}\ }\textbf {\bibinfo {volume} {87}},\
  \bibinfo {pages} {504} (\bibinfo {year} {2011})}\BibitemShut {NoStop}%
\bibitem [{\citenamefont {Ramberg}\ and\ \citenamefont
  {Stephansson}(1964)}]{ramberg1964compression}%
  \BibitemOpen
  \bibfield  {author} {\bibinfo {author} {\bibfnamefont {H.}~\bibnamefont
  {Ramberg}}\ and\ \bibinfo {author} {\bibfnamefont {O.}~\bibnamefont
  {Stephansson}},\ }\bibfield  {title} {\bibinfo {title} {Compression of
  floating elastic and viscous plates affected by gravity, a basis for
  discussing crustal buckling},\ }\href@noop {} {\bibfield  {journal} {\bibinfo
   {journal} {Tectonophysics}\ }\textbf {\bibinfo {volume} {1}},\ \bibinfo
  {pages} {101} (\bibinfo {year} {1964})}\BibitemShut {NoStop}%
\bibitem [{\citenamefont {Schmalholz}\ \emph {et~al.}(2002)\citenamefont
  {Schmalholz}, \citenamefont {Podladchikov},\ and\ \citenamefont
  {Burg}}]{schmalholz2002control}%
  \BibitemOpen
  \bibfield  {author} {\bibinfo {author} {\bibfnamefont {S.}~\bibnamefont
  {Schmalholz}}, \bibinfo {author} {\bibfnamefont {Y.}~\bibnamefont
  {Podladchikov}},\ and\ \bibinfo {author} {\bibfnamefont {J.-P.}\ \bibnamefont
  {Burg}},\ }\bibfield  {title} {\bibinfo {title} {Control of folding by
  gravity and matrix thickness: Implications for large-scale folding},\
  }\href@noop {} {\bibfield  {journal} {\bibinfo  {journal} {Journal of
  Geophysical Research: Solid Earth}\ }\textbf {\bibinfo {volume} {107}},\
  \bibinfo {pages} {ETG} (\bibinfo {year} {2002})}\BibitemShut {NoStop}%
\bibitem [{\citenamefont {Shulaker}\ \emph {et~al.}(2011)\citenamefont
  {Shulaker}, \citenamefont {Wei}, \citenamefont {Patil}, \citenamefont
  {Provine}, \citenamefont {Chen}, \citenamefont {Wong},\ and\ \citenamefont
  {Mitra}}]{shulaker2011linear}%
  \BibitemOpen
  \bibfield  {author} {\bibinfo {author} {\bibfnamefont {M.~M.}\ \bibnamefont
  {Shulaker}}, \bibinfo {author} {\bibfnamefont {H.}~\bibnamefont {Wei}},
  \bibinfo {author} {\bibfnamefont {N.}~\bibnamefont {Patil}}, \bibinfo
  {author} {\bibfnamefont {J.}~\bibnamefont {Provine}}, \bibinfo {author}
  {\bibfnamefont {H.-Y.}\ \bibnamefont {Chen}}, \bibinfo {author}
  {\bibfnamefont {H.-S.}\ \bibnamefont {Wong}},\ and\ \bibinfo {author}
  {\bibfnamefont {S.}~\bibnamefont {Mitra}},\ }\bibfield  {title} {\bibinfo
  {title} {Linear increases in carbon nanotube density through multiple
  transfer technique},\ }\href@noop {} {\bibfield  {journal} {\bibinfo
  {journal} {Nano letters}\ }\textbf {\bibinfo {volume} {11}},\ \bibinfo
  {pages} {1881} (\bibinfo {year} {2011})}\BibitemShut {NoStop}%
\bibitem [{\citenamefont {Reis}\ \emph {et~al.}(2009)\citenamefont {Reis},
  \citenamefont {Corson}, \citenamefont {Boudaoud},\ and\ \citenamefont
  {Roman}}]{reis2009localization}%
  \BibitemOpen
  \bibfield  {author} {\bibinfo {author} {\bibfnamefont {P.~M.}\ \bibnamefont
  {Reis}}, \bibinfo {author} {\bibfnamefont {F.}~\bibnamefont {Corson}},
  \bibinfo {author} {\bibfnamefont {A.}~\bibnamefont {Boudaoud}},\ and\
  \bibinfo {author} {\bibfnamefont {B.}~\bibnamefont {Roman}},\ }\bibfield
  {title} {\bibinfo {title} {Localization through surface folding in solid
  foams under compression},\ }\href@noop {} {\bibfield  {journal} {\bibinfo
  {journal} {Physical review letters}\ }\textbf {\bibinfo {volume} {103}},\
  \bibinfo {pages} {045501} (\bibinfo {year} {2009})}\BibitemShut {NoStop}%
\bibitem [{\citenamefont {Vella}\ \emph
  {et~al.}(2009{\natexlab{b}})\citenamefont {Vella}, \citenamefont {Bico},
  \citenamefont {Boudaoud}, \citenamefont {Roman},\ and\ \citenamefont
  {Reis}}]{vella2009macroscopic}%
  \BibitemOpen
  \bibfield  {author} {\bibinfo {author} {\bibfnamefont {D.}~\bibnamefont
  {Vella}}, \bibinfo {author} {\bibfnamefont {J.}~\bibnamefont {Bico}},
  \bibinfo {author} {\bibfnamefont {A.}~\bibnamefont {Boudaoud}}, \bibinfo
  {author} {\bibfnamefont {B.}~\bibnamefont {Roman}},\ and\ \bibinfo {author}
  {\bibfnamefont {P.~M.}\ \bibnamefont {Reis}},\ }\bibfield  {title} {\bibinfo
  {title} {The macroscopic delamination of thin films from elastic
  substrates},\ }\href@noop {} {\bibfield  {journal} {\bibinfo  {journal}
  {Proceedings of the National Academy of Sciences}\ }\textbf {\bibinfo
  {volume} {106}},\ \bibinfo {pages} {10901} (\bibinfo {year}
  {2009}{\natexlab{b}})}\BibitemShut {NoStop}%
\bibitem [{\citenamefont {Schwarz}\ and\ \citenamefont
  {Safran}(2013)}]{schwarz2013physics}%
  \BibitemOpen
  \bibfield  {author} {\bibinfo {author} {\bibfnamefont {U.~S.}\ \bibnamefont
  {Schwarz}}\ and\ \bibinfo {author} {\bibfnamefont {S.~A.}\ \bibnamefont
  {Safran}},\ }\bibfield  {title} {\bibinfo {title} {Physics of adherent
  cells},\ }\href@noop {} {\bibfield  {journal} {\bibinfo  {journal} {Reviews
  of Modern Physics}\ }\textbf {\bibinfo {volume} {85}},\ \bibinfo {pages}
  {1327} (\bibinfo {year} {2013})}\BibitemShut {NoStop}%
\bibitem [{\citenamefont {Ziebert}\ and\ \citenamefont
  {Aranson}(2016)}]{ziebert2016computational}%
  \BibitemOpen
  \bibfield  {author} {\bibinfo {author} {\bibfnamefont {F.}~\bibnamefont
  {Ziebert}}\ and\ \bibinfo {author} {\bibfnamefont {I.~S.}\ \bibnamefont
  {Aranson}},\ }\bibfield  {title} {\bibinfo {title} {Computational approaches
  to substrate-based cell motility},\ }\href@noop {} {\bibfield  {journal}
  {\bibinfo  {journal} {npj Computational Materials}\ }\textbf {\bibinfo
  {volume} {2}},\ \bibinfo {pages} {16019} (\bibinfo {year}
  {2016})}\BibitemShut {NoStop}%
\bibitem [{\citenamefont {Trueman}(1975)}]{trueman1975locomotion}%
  \BibitemOpen
  \bibfield  {author} {\bibinfo {author} {\bibfnamefont {E.~R.}\ \bibnamefont
  {Trueman}},\ }\href@noop {} {\emph {\bibinfo {title} {Locomotion of
  soft-bodied animals}}}\ (\bibinfo  {publisher} {Edward Arnold},\ \bibinfo
  {year} {1975})\BibitemShut {NoStop}%
\bibitem [{\citenamefont {Cohen}\ \emph {et~al.}(2018)\citenamefont {Cohen},
  \citenamefont {Chan},\ and\ \citenamefont {Mahadevan}}]{cohen2018competing}%
  \BibitemOpen
  \bibfield  {author} {\bibinfo {author} {\bibfnamefont {T.}~\bibnamefont
  {Cohen}}, \bibinfo {author} {\bibfnamefont {C.~U.}\ \bibnamefont {Chan}},\
  and\ \bibinfo {author} {\bibfnamefont {L.}~\bibnamefont {Mahadevan}},\
  }\bibfield  {title} {\bibinfo {title} {Competing failure modes in finite
  adhesive pads},\ }\href@noop {} {\bibfield  {journal} {\bibinfo  {journal}
  {Soft matter}\ }\textbf {\bibinfo {volume} {14}},\ \bibinfo {pages} {1771}
  (\bibinfo {year} {2018})}\BibitemShut {NoStop}%
\bibitem [{\citenamefont {Labonte}\ \emph {et~al.}(2016)\citenamefont
  {Labonte}, \citenamefont {Clemente}, \citenamefont {Dittrich}, \citenamefont
  {Kuo}, \citenamefont {Crosby}, \citenamefont {Irschick},\ and\ \citenamefont
  {Federle}}]{labonte2016extreme}%
  \BibitemOpen
  \bibfield  {author} {\bibinfo {author} {\bibfnamefont {D.}~\bibnamefont
  {Labonte}}, \bibinfo {author} {\bibfnamefont {C.~J.}\ \bibnamefont
  {Clemente}}, \bibinfo {author} {\bibfnamefont {A.}~\bibnamefont {Dittrich}},
  \bibinfo {author} {\bibfnamefont {C.-Y.}\ \bibnamefont {Kuo}}, \bibinfo
  {author} {\bibfnamefont {A.~J.}\ \bibnamefont {Crosby}}, \bibinfo {author}
  {\bibfnamefont {D.~J.}\ \bibnamefont {Irschick}},\ and\ \bibinfo {author}
  {\bibfnamefont {W.}~\bibnamefont {Federle}},\ }\bibfield  {title} {\bibinfo
  {title} {Extreme positive allometry of animal adhesive pads and the size
  limits of adhesion-based climbing},\ }\href@noop {} {\bibfield  {journal}
  {\bibinfo  {journal} {Proceedings of the National Academy of Sciences}\
  }\textbf {\bibinfo {volume} {113}},\ \bibinfo {pages} {1297} (\bibinfo {year}
  {2016})}\BibitemShut {NoStop}%
\bibitem [{SI()}]{SI}%
  \BibitemOpen
  \bibfield  {title} {\bibinfo {title} {See supplemental material at [url] for
  video.},\ }\href@noop {} {\ }\BibitemShut {NoStop}%
\bibitem [{Note1()}]{Note1}%
  \BibitemOpen
  \bibinfo {note} {It has been verified that displacements due to compression
  of the delaminated region are negligible.}\BibitemShut {Stop}%
\bibitem [{\citenamefont {Sofla}\ \emph {et~al.}(2010)\citenamefont {Sofla},
  \citenamefont {Seker}, \citenamefont {Landers},\ and\ \citenamefont
  {Begley}}]{sofla2010pdms}%
  \BibitemOpen
  \bibfield  {author} {\bibinfo {author} {\bibfnamefont {A.}~\bibnamefont
  {Sofla}}, \bibinfo {author} {\bibfnamefont {E.}~\bibnamefont {Seker}},
  \bibinfo {author} {\bibfnamefont {J.~P.}\ \bibnamefont {Landers}},\ and\
  \bibinfo {author} {\bibfnamefont {M.~R.}\ \bibnamefont {Begley}},\ }\bibfield
   {title} {\bibinfo {title} {Pdms-glass interface adhesion energy determined
  via comprehensive solutions for thin film bulge/blister tests},\ }\href@noop
  {} {\bibfield  {journal} {\bibinfo  {journal} {Journal of Applied Mechanics}\
  }\textbf {\bibinfo {volume} {77}} (\bibinfo {year} {2010})}\BibitemShut
  {NoStop}%
\bibitem [{Note2()}]{Note2}%
  \BibitemOpen
  \bibinfo {note} {Note that in PDMS surface properties vary significantly with
  stiffness.}\BibitemShut {Stop}%
\bibitem [{\citenamefont {Lin}\ \emph {et~al.}(2008)\citenamefont {Lin},
  \citenamefont {Vajpayee}, \citenamefont {Jagota}, \citenamefont {Hui},\ and\
  \citenamefont {Yang}}]{lin2008mechanically}%
  \BibitemOpen
  \bibfield  {author} {\bibinfo {author} {\bibfnamefont {P.-C.}\ \bibnamefont
  {Lin}}, \bibinfo {author} {\bibfnamefont {S.}~\bibnamefont {Vajpayee}},
  \bibinfo {author} {\bibfnamefont {A.}~\bibnamefont {Jagota}}, \bibinfo
  {author} {\bibfnamefont {C.-Y.}\ \bibnamefont {Hui}},\ and\ \bibinfo {author}
  {\bibfnamefont {S.}~\bibnamefont {Yang}},\ }\bibfield  {title} {\bibinfo
  {title} {Mechanically tunable dry adhesive from wrinkled elastomers},\
  }\href@noop {} {\bibfield  {journal} {\bibinfo  {journal} {Soft Matter}\
  }\textbf {\bibinfo {volume} {4}},\ \bibinfo {pages} {1830} (\bibinfo {year}
  {2008})}\BibitemShut {NoStop}%
\bibitem [{\citenamefont {Hwang}\ \emph {et~al.}(2018)\citenamefont {Hwang},
  \citenamefont {Trent},\ and\ \citenamefont {Bartlett}}]{hwang2018kirigami}%
  \BibitemOpen
  \bibfield  {author} {\bibinfo {author} {\bibfnamefont {D.-G.}\ \bibnamefont
  {Hwang}}, \bibinfo {author} {\bibfnamefont {K.}~\bibnamefont {Trent}},\ and\
  \bibinfo {author} {\bibfnamefont {M.~D.}\ \bibnamefont {Bartlett}},\
  }\bibfield  {title} {\bibinfo {title} {Kirigami-inspired structures for smart
  adhesion},\ }\href@noop {} {\bibfield  {journal} {\bibinfo  {journal} {ACS
  applied materials \& interfaces}\ }\textbf {\bibinfo {volume} {10}},\
  \bibinfo {pages} {6747} (\bibinfo {year} {2018})}\BibitemShut {NoStop}%
\end{thebibliography}
\end{document}